\documentclass[12pt]{article}

\usepackage{graphicx}
\usepackage{amsmath}
\usepackage{amssymb}
\usepackage{bigcircle}
\usepackage{wrapfig}
\usepackage{indentfirst}

\begin{document}

\vskip 0.5cm \centerline{\bf\Large Deconvoluting GPD from an explicit DVCS amplitude}  \vskip 0.3cm
\centerline{K.~Bondarenko $^{a\diamond}$ and L.~Jenkovszky $^{b\star}$}

\vskip 1cm

\centerline{$^a$ \sl Taras Shevchenko National University of Kyiv} \centerline{$^b$ \sl Bogolubov Institute for Theoretical Physics,
National Academy of Sciences of Ukraine} \centerline{\sl Kiev,
03680 Ukraine} \vskip
0.1cm

\begin{abstract}\noindent
By assuming factorization of the GPD under the deconvolution integral for the hand-bag diagram,
we develop a method of solving this integral beyond the cross-over line.
As examples we use explicit models of deeply virtual Compton scattering (DVCS)
amplitudes to get solution for relevant GPDs.

\end{abstract}

\vskip 0.1cm

$
\begin{array}{ll}

^{\diamond}\mbox{{\it e-mail address:}} &
   \mbox{kyrylo.bondarenko@gmail.com} \\
^{\star}\mbox{{\it e-mail address:}} &
   \mbox{jenk@bitp.kiev.ua} \\

%^{\ast}\mbox{{\it e-mail address:}} &
 %  \mbox{?@?.?} \\

\end{array}
$

%\end{titlepage}

%\section{Convolution equation}\label{s1}
\section{Introduction}
Deeply virtual Compton scattering (DVCS) combines the features of inelastic processes
with those of an elastic one. It has been realized \cite{1, 2, Ji} that a straightforward generalization of the ordinary parton densities arises in exclusive two-photon processes in the Bjorken region, e.g. in Compton scattering with a highly virtual incoming photon, and in hard photoproduction of mesons.

\begin{figure}[h]
\begin{center}
\includegraphics[clip,width=6in]{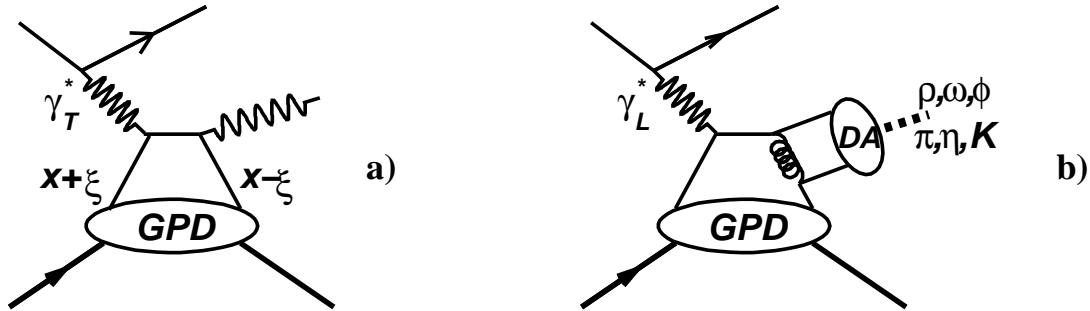}
\end{center}
\vspace{-0.8cm} \caption{\small\it{ a) Hand-bag diagram for DVCS; b) Hand-bag diagram for vector meson production}}
\label{fig:handbag}
\end{figure}

Generalized parton distributions (GPD) combine our knowledge about the one-di\-men\-sional parton distribution in the longitudinal momentum with the impact-parameter, or transverse distribution of matter in a hadron or nucleus\cite{1, 2, Ji, Kumericki}.
GPD cannot be measured directly, instead they appear as convolution integrals of the form
\begin{equation}\label{Convolut}
A(\xi, t, Q^2)=\int\limits_{-1}^1\frac{\text{GPD}(x,\xi, t, Q^2)}{x-\xi+i\varepsilon}dx,
\end{equation}
where $\xi\approx\dfrac{x_{Bj}}{2-x_{Bj}}$ is called skewness and $x$ is the average longitudinal momentum fraction of the struck parton in the initial and final states. This integral corresponds to deeply virtual Compton scattering (DVCS) in the "hand-bag" approximation, see Fig. \ref{fig:handbag} a), and $x$ is the integration variable, not to be 
confused with the Bjorken variable $x_{Bj}$.

Eq. (\ref{Convolut}) is an integral equation for the unknown function GPD. The solution is well known along the cross-over line $x=\xi$, where (see e.g. \cite{Kumericki})
\begin{equation}
 \text{GPD}(x=\xi,\xi,t,Q^2)=-\frac{1}{\pi}\text{Im}A(\xi,t,Q^2).
\end{equation}

GPD (the nominator in Eq. (\ref{Convolut})) is assumed to be universal for all exclusive processes of the
type $\gamma^*p\rightarrow Vp,$ where $V$ stands for a real photon or vector meson, the dependence on the
processes coming from the "handle", calculable perturbatively.  If so, many different processes of exclusive
vector particle production can be described, according to Eq. (\ref{Convolut}), by a universal GPD with different integration kernels, see Fig. \ref{fig:handbag} b).
Since the GPD is not known apriori, one seeks for models of GPD (see e.g. \cite{3}) based on general constraints on its analytic and asymptotic
behavior. The calculated scattering amplitudes (cross sections) are than compared with the data to confirm, modify or reject the chosen form of the GPD.

In this paper we use two explicit models of DVCS and vector meson production (VMP) to obtain exact analytic expressions for GPD. To this end we propose a method of deconvolution of integral equations with a singular kernel of the Cauchy type to enable to go beyond the crossover trajectory and explore the hitherto unknown properties of the GPD.

Regge pole models provide an adequate framework to describe high-energy, low $t$ scattering phenomena. Being part of the $S$ matrix theory, however,
strictly speaking, they are valid only for the scattering of on-mass-shall particles. Still, the successful application of the Regge pole models in
describing the off mass-shall HERA data opened the way to their use deep-inelastic scattering, DVCS  and vector meson production at HERA. The appearance of
the "forbidden" $Q^2$ dependence was circumvent by calling the $Q^2$-dependent Regge trajectories "effective" ones. A particularly simple and efficient
Regge pole model \cite{Capua} with $Q^2$-dependent residues (vertices) will be used below to guide  our deconvolution procedure.

In a more advanced Regge-pole model \cite{Reggeometry}, to be used below, the Regge trajectories and the residues do not depend on virtuality. Instead, the
amplitude contains two (or more) Regge-pole terms, whose relative weight depends on $Q^2$, mimicking the multi-pole nature of the so-called QCD Pomeron.
Of great interest is the use of
alternative models \cite{Mullerfit}.
It is important to note that the Regge pole amplitudes are complex functions, their phase being fixed by the residue function. Since these model satisfy
the basic properties of the theory, yet they fit the data, we shall use them to demonstrate the merits of our deconvolution procedure. %Let us remind that

The paper is organized as follows: in Section 2 (Simple DVCS amplitude) we introduce an explicit DVCS amplitude whose merit it simplicity, necessary in testing the
capacity of our deconvolution procedure; in Section 3 (Deconvolution) we describe the mathematical basis for our method; in Section 4 (Factorization) we introduce the factorization hypothesis, essential in deriving GPD. In Sections 5 (Reggeometric model of DVCS) we apply our method to a more complicated model for the amplitude.

\section{A simple DVCS amplitude}

For the sake of clarity we start with a very simply but efficient model of the DVCS scattering amplitude \cite{Capua}. This model accumulate in a compact way the main
properties of the expected DVCS amplitude. Namely, it is Regge-behaved, has  the required behavior in s, t and $Q^2$, yet it fits the DVCS data measured at HERA by the
Hi and ZEUS Collaborations. The kinematics of those data are such that they correspond to diffractive scattering, and consequently the $t$ channel exchange is dominated by
a single Pomeron trajectory.

The explicit forms of the relevant scattering amplitude is
\cite{Capua}:

\begin{equation}\label{A2}
A(s,t,Q^2)_{\gamma^* p\rightarrow\gamma p}=
-A_0e^{b_1\alpha(t)}e^{b_2 \beta(z)}(-is/s_0)^{\alpha(t)}=
-A_0e^{(b_1+L)\alpha(t)+b_2\beta(z)},
\end{equation}
where $L\equiv\ln(-is/s_0)$,
\begin{equation}\label{alpha}
\alpha(t)=\alpha(0)-\alpha_1\ln(1-\alpha_2 t),
\end{equation}
and
\begin{equation}\label{beta}
\beta(z)=\alpha(0)-\alpha_1\ln(1-\alpha_2 z),
\end{equation}
where $z=t-Q^2$ is a new variable introduced in Ref. \cite{Capua}.

From Eq. (\ref{A2}) we get the real and imaginary parts of the
DVCS amplitude:
\begin{equation}
\text{Im} A_{DVCS}(x_{Bj},t,Q_0^2)=
\sin\Bigl(\frac{\pi\alpha(t)}{2}\Bigr)G(t,Q_0^2)\Bigl(\frac{Q_0^2}{s_0x_{Bj}}\Bigr)^{\alpha(t)},
\end{equation}

\begin{equation}
\text{Re} A_{DVCS}(x_{Bj},t,Q_0^2)=
-\cos\Bigl(\frac{\pi\alpha(t)}{2}\Bigr)G(t,Q_0^2)\Bigl(\frac{Q_0^2}{s_0x_{Bj}}\Bigr)^{\alpha(t)},
\end{equation}
where
\begin{equation}
G(t,Q^2_0)=e^{b(\alpha(t)+\beta(t,Q^2_0))}.
\end{equation}
Skewness is defined in terms of the Bjorken variable $x_{Bj}$ as
$\xi\simeq \frac{x_{Bj}}{2-x_{Bj}}$ or v.v., $x_{Bj}\simeq \frac{2\xi}{1+\xi}$.

Since the $Q^2$ dependence in the model of Ref. \cite{Capua} may not follow QCD evolution (ambiguous in DVCS),
we keep it fixed at some value $Q^2$ that may be associated with the "frozen" QCD coupling constant. It has no
effect on the deconvolution procedure in question.

In the deconvolution procedure, Eq. (\ref{Convolut}), the variables $t$ and $Q^2$ appear merely as
"parameters", therefore, for simplicity, we rewrite $A_{DVCS}$ as:
\begin{equation}\label{short_form_A}
A_{DVCS}(\xi,t,Q^2)=-e^{-\frac{i\pi\alpha(t)}{2}}B(t,Q^2)\left(\frac{1+\xi}{\xi}\right)^{\alpha(t)},
\end{equation}
where
\begin{equation}\label{B_function}
B(t,Q^2)=G(t,Q_0^2)\Bigl(\frac{Q_0^2}{2s_0}\Bigr)^{\alpha(t)}
\end{equation}
is a real-valued function.

With these ingredients we can now proceed to the deconvolution procedure.

\section{Deconvolution}
By substituting Eq. (\ref{short_form_A}) in Eq. (\ref{Convolut}) we get (the dependence on the "parameters" is suppressed):
\begin{equation}
	 -e^{-\frac{i\pi\alpha}{2}}B\left(\frac{1+\xi}{\xi}\right)^{\alpha}=\int\limits_{-1}^{1}\frac{\text{GPD}(x,\xi)}{x-\xi+i\varepsilon}dx,
\end{equation}
where $\alpha<1$ and $B$ is defined by Eq. (\ref{B_function}). This is a singular integral equation with a Cauchy-type unknown kernel. By using the Sokhotski rule ($\frac{1}{x+i\varepsilon}=\text{p.v.}\left(\frac{1}{x}\right)-i\pi\delta(x)$), we obtain:
\begin{equation}
	 -e^{-\frac{i\pi\alpha}{2}}B\left(\frac{1+\xi}{\xi}\right)^{\alpha}=\text{p.v.}\int\limits^{1}_{-1}\frac{\text{GPD}(x,\xi)}{x-\xi}dx-i\pi \text{GPD}(\xi,\xi).
\end{equation}

Since GPD is a real-valued function, we obtain its important property, called the cross-over trajectory condition:
\begin{equation}\label{cross-over}
	\text{GPD}(x=\xi,\xi,t,Q^2)=-\frac{1}{\pi}\text{Im}A_{DVCS}(\xi,t,Q^2).
\end{equation}

At this moment we need to make approximation to proceed. Assume that GPD function could be factorized
\begin{equation}
	\label{factorization}
	\text{GPD}(x,\xi)=f(\xi)\varphi(x),
\end{equation}
we get
\begin{equation}
	\label{MyEq3}
	 -\frac{e^{-\frac{i\pi\alpha}{2}}B}{f(\xi)}\left(\frac{1+\xi}{\xi}\right)^{\alpha}=\text{p.v.}\int\limits_{-1}^{1}\frac{\varphi(x)}{x-\xi}dx-i\pi \varphi(\xi).
\end{equation}
This is an integral equation for $\phi(x)$ to be resolved. As we know from the Sokhotski-Plemelj theorem \cite{PolyaninManzhirov}, there exist functions $\Phi^{+}(\xi)$ and $\Phi^{-}(\xi)$ in the complex plain of the $\xi$ variable such that: $\Phi^{+}(\xi)$ is defined for $\text{Im}(\xi)\geq 0$ and is analytic for $\text{Im}(\xi)>0$; $\Phi^{-}(\xi)$ is defined for $\text{Im}(\xi)\leq 0$ and is analytic for $\text{Im}(\xi)<0$, $\Phi^{+}(\xi)$, while $\Phi^{-}(\xi)$ are continuous on $\mathbb{R}$ (except fir a few points of the integrable discontinuity), and
\begin{equation}
	\label{MyEq1}
	\Phi^{+}(\xi)+\Phi^{-}(\xi)=\frac{1}{\pi i}\text{p.v.}\int\limits_{-1}^{1}\frac{\varphi(x)}{x-\xi}dx,\hphantom{\varphi(x)}\xi\in (-1,1),\hphantom{\mathbb{R}\setminus [-1,1]}
\end{equation}
\begin{equation}
	\label{MyEq2}
	\Phi^{+}(x)-\Phi^{-}(x)=\varphi(x),\hphantom{\frac{1}{\pi i}\text{p.v.}\int\limits_{-1}^{1}\frac{\varphi(x)}{x-\xi}dx}x\in (-1,1),\hphantom{\mathbb{R}\setminus [-1,1]}
\end{equation}
\begin{equation}
	\label{MyEq4}
	\Phi^{+}(\xi)=\Phi^{-}(\xi),\hphantom{-\varphi(x),\frac{1}{\pi i}\text{p.v.}\int\limits_{-1}^{1}\frac{\varphi(x)}{x-\xi}dx}\xi\in\mathbb{R}\setminus [-1,1].\hphantom{(-1,1)}
\end{equation}
Here Eq. (\ref{MyEq4}) is a self-consistency condition for open-loop contours, see \cite{PolyaninManzhirov}, Sec. 14.3-11 (for a
 more profound treatment of the problem see  \cite{Gakhov}). This condition means (Eq. (\ref{MyEq2})) that $\varphi(x)=0$ for $x\in\mathbb{R}\setminus [-1,1]$. An alternative interpretation of this condition we will discussed below.

By substituting Eqs. (\ref{MyEq1}) and (\ref{MyEq2}) into Eq. (\ref{MyEq3},) we arrive at the Riemann problem:
\begin{equation}
	\label{MyEq5}
	\Phi^{-}(\xi)=-\frac{e^{-\frac{i\pi\alpha}{2}}B}{2\pi i f(\xi)}\left(\frac{1+\xi}{\xi}\right)^{\alpha},\text{\quad}\xi\in (-1,1).
\end{equation}

The analytic continuation of $\Phi^{-}(\xi)$ for $\text{Im}(\xi)<0$ if very simple: $\Phi^{-}(\xi)=\frac{e^{-\frac{i\pi\alpha}{2}}B}{2\pi i f(\xi)}\left(\frac{1+\xi}{\xi}\right)^{\alpha}$. From Eq. (\ref{MyEq4}) we can also find $\Phi^{+}(\xi)$ as an analytic continuation of $\Phi^{-}(\xi)$:
\begin{equation}
	\label{MyEq6}
	\Phi^{+}(\xi)=-\frac{e^{-\frac{i\pi\alpha}{2}}B}{2\pi i f(\xi)}\left(\frac{1+\xi}{\xi}\right)^{\alpha},\text{\quad}\text{Im}(\xi)>0.
\end{equation}

Suppose $f(\xi)$ has no zeros or poles within the interval $(-1,1)$. In this case we can write the solution even without any knowledge of the exact form of $f(\xi)$. The function $\Phi^{+}(\xi)$ has an order $\alpha$ zero at the point $-1$ and an order $\alpha$ pole at the point $0$. The cut from $-1$ to $0$ along $\mathbb{R}$ and Eq. (\ref{MyEq4}) almost define $\Phi^{+}(\xi)$ on $(-1,1)\setminus\{0\}$:
\begin{equation}
	\label{MyEq7}
	\Phi^{+}(\xi)=-e^{-2\pi i\alpha}\frac{e^{-\frac{i\pi\alpha}{2}}B}{2\pi i f(\xi)}\left(\frac{1+\xi}{\xi}\right)^{\alpha},\text{\quad}\xi\in (-1,0),
\end{equation}
\begin{equation}
	\label{MyEq10}
	\Phi^{+}(\xi)=-\frac{e^{-\frac{i\pi\alpha}{2}}B}{2\pi i f(\xi)}\left(\frac{1+\xi}{\xi}\right)^{\alpha},\hphantom{e^{-2\pi i\alpha}}\text{\quad}\xi\in (0,1).\hphantom{-}
\end{equation}

\thicklines\unitlength=1cm
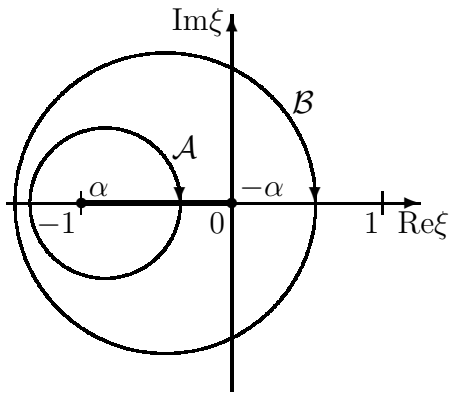
\begin{wrapfigure}{l}{6cm}
	\begin{picture}(6,6)
		\put(3,0.5){\vector(0,1){5}}
		\put(0.0,3){\line(1,0){1.0}}
		\put(3,3){\vector(1,0){2.5}}
		\put(2.2,5.3){$\text{Im}\xi$}
		\put(5.2,2.6){$\text{Re}\xi$}
		\put(2.7,2.6){$0$}
		\put(3.0,3.0){\circle*{0.15}}
		\put(3.1,3.1){$-\alpha$}
		\put(1.0,3.0){\circle*{0.15}}
		\put(1.1,3.1){$\alpha$}
		\thinlines
		\put(0.99,2.85){\line(0,1){0.3}}
		\put(0.4,2.6){$-1$}
		\put(5.0,2.85){\line(0,1){0.3}}
		\put(4.75,2.6){$1$}
		\linethickness{0.7mm}
		\put(1.0,3.0){\line(1,0){2}}
		\thicklines
		\bigcircle{1.3}{3.0}{1.0}
		\put(2.3,3.0){\vector(0,-1){0}}
		\bigcircle{2.1}{3.0}{2.0}
		\put(4.1,3.0){\vector(0,-1){0}}
		\put(2.2,3.6){$\mathcal{A}$}
		\put(3.8,4.2){$\mathcal{B}$}
	\end{picture}
	\caption{\small\it The function $\Phi$ on the complex $\xi$ plane}
	\label{fig:complexplane}
\end{wrapfigure}

For a better understanding of this result, look at this problem from a different perspective: Eq. (\ref{MyEq4}) may be interpret as if there existed only one complex function on the $\xi$ plane --- $\Phi$, while $\Phi^+$ and $\Phi^-$ are values of this function for $\text{Im}\xi>0$ and $\text{Im}\xi<0$, respectively. $\Phi$ is a holomorphic function with branching points at $-1$ and $0$, where it has an order $\alpha$ zero and an order $\alpha$ pole. To determine its values unambiguously on all complex plane, we cut the plane from $-1$ to $0$ (see Fig. (\ref{fig:complexplane})). By encircling the point $-1$, initiating in the segment $(-1;0)$ (contour $\mathcal{A}$ in the figure), the phase of function $\Phi$ will change by a factor $e^{-2\pi i\alpha}$. However, if starting point will be in the segment $(0;1)$ (contour $\mathcal{B}$ on the figure), the phase of $\Phi$ will not change since the factors coming from the zero and the pole will cancel; each other.

From Eq. (\ref{MyEq2}) we have:
\begin{equation}
	\label{MyEq8}
	\varphi(x)=\left(1-e^{-2\pi i\alpha}\right)\frac{e^{-\frac{i\pi\alpha}{2}}B}{2\pi i f(x)}\left(\frac{1+x}{x}\right)^{\alpha},\hphantom{0}x\in (-1,0),
\end{equation}
\begin{equation}
	\label{MyEq11}
	\varphi(x)=0,\hphantom{1-\left(e^{-2\pi i\alpha}\right)\frac{e^{-\frac{i\pi\alpha}{2}}B}{2\pi i f(x)}\left(\frac{1+x}{x}\right)^{\alpha}}x\in (0,1).\hphantom{-}
\end{equation}

Finally, for the GPD we obtain the solution:
\begin{equation}
	\label{MyEq9}
	\text{GPD}(x,\xi)=\left(1-e^{-2\pi i\alpha}\right)\frac{e^{-\frac{i\pi\alpha}{2}}B f(\xi)}{2\pi i f(x)}\left(\frac{1+x}{x}\right)^{\alpha},\hphantom{0}x\in (-1,0),
\end{equation}
\begin{equation}
	\label{MyEq12}
	\text{GPD}(x,\xi)=0,\hphantom{1-\left(e^{-2\pi i\alpha}\right)\frac{e^{-\frac{i\pi\alpha}{2}}B f(\xi)}{2\pi i f(x)}\left(\frac{1+x}{x}\right)^{\alpha}}x\in (0,1).\hphantom{-}
\end{equation}

We see that this solution is unphysical, because GPD must be a real-valued function for physical values of $\xi$ ($0<\xi<1$, see \cite{Ji}). In the next section we discuss how to satisfy this condition.

\section{Factorization}

In the previous section we have used the factorization assumption for the deconvolution of Eq. (\ref{Convolut}). At this point it is hard to say how good is this assumption. We shall estimate it by its consequences.

As we have seen in the previous section, the solution for GPD critically depends on the analytic properties of the left-hand side of the Eq. (\ref{MyEq3}) for $\xi\in(-1;1)$. Our degree of freedom is the function $f(\xi)$ in Eq. (\ref{factorization}). Its zeros and poles may help us in getting physical values for GPD.

Let us construct a solution that will be a real-valued function for $x\in (a,b)$, where $(a,b)$ is an arbitrary segment along the real axis. It is easy to check, that the function
\begin{equation}\label{goodf}
	 f(\xi)=\left(\frac{1+\xi}{\xi}\right)^{\alpha}(\xi-a)^{\frac{\alpha}{2}}(b-\xi)^{-\frac{\alpha}{2}}g(\xi),
\end{equation}
solves this problem. Here $g(\xi)$ is an arbitrary real-valued function without zeros or discontinuities of fractional order along $[-1,1]$. Let us check this.

By substituting (\ref{goodf}) into (\ref{MyEq5}), we get:
\begin{equation}
	\Phi^{-}(\xi)=-\frac{e^{-\frac{i\pi\alpha}{2}}B}{2\pi i g(\xi)}(\xi-a)^{-\frac{\alpha}{2}}(b-\xi)^{\frac{\alpha}{2}},\text{\quad}\xi\in (-1,1).
\end{equation}

By repeating the procedure from the previous section, we see that the expression for $\Phi^+$ ($g(\xi)$ has only zeros or discontinuities of integer order, so it changes phase only by integer number of $2\pi i$):
\begin{equation}
	\Phi^{+}(\xi)=-\frac{e^{\frac{i\pi\alpha}{2}}B}{2\pi i g(\xi)}(\xi-a)^{-\frac{\alpha}{2}}(b-\xi)^{\frac{\alpha}{2}},\text{\quad}\xi\in (a,b),\hphantom{\mathbb{R}\setminus [a,b]}
\end{equation}
\begin{equation}
	\Phi^{+}(\xi)=-\frac{e^{-\frac{i\pi\alpha}{2}}B}{2\pi i g(\xi)}(\xi-a)^{-\frac{\alpha}{2}}(b-\xi)^{\frac{\alpha}{2}},\text{\quad}\xi\in \mathbb{R}\setminus [a,b].\hphantom{(a,b)}
\end{equation}

From Eq. (\ref{MyEq2}) we have:
\begin{equation}
	\varphi(x)=-\sin\left(\frac{\pi\alpha}{2}\right)\frac{B}{\pi g(x)}(x-a)^{-\frac{\alpha}{2}}(b-x)^{\frac{\alpha}{2}},\hphantom{0}x\in (a,b),\hphantom{\mathbb{R}\setminus[a,b]}
\end{equation}
\begin{equation}
	\varphi(x)=0,\hphantom{-\sin\left(\frac{\pi\alpha}{2}\right)\frac{B}{\pi g(x)}(x-a)^{-\frac{\alpha}{2}}(b-x)^{\frac{\alpha}{2}}}x\in \mathbb{R}\setminus[a,b].\hphantom{(a,b)}
\end{equation}

Accordingly, for GPD we have:
\begin{equation}
	 \text{GPD}(x,\xi)=-\sin\left(\frac{\pi\alpha}{2}\right)\frac{B}{\pi}\frac{g(\xi)}{g(x)}\left(\frac{(\xi-a)(b-x)}{(x-a)(b-\xi)}\right)^{\frac{\alpha}{2}}\left(\frac{1+\xi}{\xi}\right)^{\alpha},\hphantom{0}x\in (a,b),\hphantom{\mathbb{R}\setminus[a,b]}
\end{equation}
\begin{equation}
	 \text{GPD}(x,\xi)=0,\hphantom{-\sin\left(\frac{\pi\alpha}{2}\right)\frac{B}{\pi}\frac{g(\xi)}{g(x)}\left(\frac{(\xi-a)(b-x)}{(x-a)(b-\xi)}\right)^{\frac{\alpha}{2}}\left(\frac{1+\xi}{\xi}\right)^{\alpha}}x\in \mathbb{R}\setminus[a,b].\hphantom{(a,b)}
\end{equation}

The most interesting case for us is that of $a=-1$, $b=1$, then GPD is given by the expression:
\begin{multline}\label{solGPD}
	 \text{GPD}(x,\xi,t,Q^2)=-\sin\left(\frac{\pi\alpha(t)}{2}\right)\frac{B(t,Q^2)}{\pi}\frac{g(\xi)}{g(x)}\left(\frac{(1+\xi)(1-x)}{(1+x)(1-\xi)}\right)^{\frac{\alpha(t)}{2}}\left(\frac{1+\xi}{\xi}\right)^{\alpha(t)}=\\=-\frac{1}{\pi}\frac{g(\xi)}{g(x)}\left(\frac{(1+\xi)(1-x)}{(1+x)(1-\xi)}\right)^{\frac{\alpha(t)}{2}}\text{Im}A_{DVCS}(\xi,t,Q^2).
\end{multline}

This GPD satisfies the cross-over trajectory condition (\ref{cross-over}). It is easy to see that any other function $f$ except (\ref{goodf}) will violate the reality of GPD. So, this is a general solution.

\section{Reggeometric model of DVCS}
Another simple and explicit albeit more advanced model for DVCS, as well as for vector meson production (VMP) was proposed recently \cite{Reggeometry}.

Here the Pomeron is also unique for all reactions, but the scattering amplitude contains two terms, a "soft" ($s$) and "hard" ($h$) one, "weighted" by $\widetilde{Q^2}$-dependent pre-factors:
\begin{multline}\label{Amplitude2}
	 A(s,t,Q^2,M_v^2)=\frac{\tilde{A_s}}{\Bigl(1+\frac{\widetilde{Q^2}}{\widetilde{Q_s^2}}\Bigr)^{n_s}}e^{-i\frac{\pi}{2}\alpha_s(t)}\Bigl(\frac{s}{s_{0s}}\Bigr)^{\alpha_s(t)} e^{2\Bigl(\frac{a_s}{\widetilde{Q^2}}+\frac{b_s}{2m_p^2}\Bigr)t}+\\+\frac{\tilde{A_h}\Bigl(\frac{\widetilde{Q^2}}{\widetilde{Q_h^2}}\Bigr)}{{\Bigl(1+\frac{\widetilde{Q^2}}{\widetilde{Q_h^2}}\Bigr)}^{n_h+1}}e^{-i\frac{\pi}{2}\alpha_h(t)}\Bigl(\frac{s}{s_{0h}}\Bigr)^{\alpha_h(t)} e^{2\Bigl(\frac{a_h}{\widetilde{Q^2}}+\frac{b_h}{2m_p^2}\Bigr)t}.
\end{multline}

In a sense, it mimics the multi(infinite)-component QCD Pomeron and, in principle, is applicable to any exclusive reactions, be it "soft" or "hard".
The parameters of the model are fixed from "first principles" and/or from the fits to the data, see \cite{Reggeometry}.

Similarly to Sec. 3, and by using the relation $s=\frac{Q^2(1+\xi)}{2\xi}$, we rewrite this amplitude in terms of the variable characteristic of GPD:
\begin{equation}\label{short_form_A2}
	 A(\xi,t,Q^2,M_v^2)=e^{-\frac{i\pi\alpha_s(t)}{2}}B_s(t,Q^2,M_v^2)\left(\frac{1+\xi}{\xi}\right)^{\alpha_s(t)}+e^{-\frac{i\pi\alpha_h(t)}{2}}B_h(t,Q^2,M_v^2)\left(\frac{1+\xi}{\xi}\right)^{\alpha_h(t)},
\end{equation}
where
\begin{equation}
	 B_i(t,Q^2,M_v^2)=\frac{\tilde{A_i}}{\Bigl(1+\frac{\widetilde{Q^2}}{\widetilde{Q_i^2}}\Bigr)^{n_i}}\Bigl(\frac{Q^2}{2s_{0i}}\Bigr)^{\alpha_i(t)} e^{2\Bigl(\frac{a_i}{\widetilde{Q^2}}+\frac{b_i}{2m_p^2}\Bigr)t},\text{\quad}i=s,h.
\end{equation}

It is easy to see that our simple factorization approach is not working here any more. Really, let assume factorization to be of the form $$\text{GPD}(x,\xi,t,Q^2,M_v^2)=f(\xi,t,Q^2,M_v^2)\varphi(x,t,Q^2,M_v^2),$$ with an arbitrary function $f$. The relevant solution will appear in the form (the dependence on the "parameters" $t,Q^2,M_v^2$ is again hidden):
\begin{multline}\label{BreakReality}
	\text{GPD}(x,\xi)=\frac{f(\xi)}{2\pi i f(x)}\left(\left(1-e^{-2\pi i(\alpha_s+\alpha_f(\xi))}\right)e^{-\frac{i\pi\alpha_s}{2}}B_s\left(\frac{1+x}{x}\right)^{\alpha_s}+\right.\\+\left. \left(1-e^{-2\pi i(\alpha_h+\alpha_f(\xi))}\right)e^{-\frac{i\pi\alpha_h}{2}}B_h\left(\frac{1+x}{x}\right)^{\alpha_h}\right),
\end{multline}
where $\alpha_f(\xi)$ is an additive phase shift depending on the analytic properties of $f$. Let $\alpha_s\ne \alpha_h$ and their values be independent. Then we can arrange the GPD to be a real-valued function for a certain point $x=x_0,\xi=x_0$, both terms in (\ref{BreakReality}) complex, their being real. However, by changing the value of $x$, we break this condition, since $\alpha_f$ cannot depend on $x$.

We may proceed in the following way: the equation for GPD (\ref{Convolut}) is linear, so we can split our amplitude (\ref{short_form_A2}) in two parts and search for the solution as the sum of solutions for both parts.

Let $A(\xi,t,Q^2,M_v^2)=A_s(\xi,t,Q^2,M_v^2)+A_h(\xi,t,Q^2,M_v^2)$, where:
\begin{equation}
	 A_s(\xi,t,Q^2,M_v^2)=e^{-\frac{i\pi\alpha_s(t)}{2}}B_s(t,Q^2,M_v^2)\left(\frac{1+\xi}{\xi}\right)^{\alpha_s(t)},
\end{equation}
\begin{equation}
	 A_h(\xi,t,Q^2,M_v^2)=e^{-\frac{i\pi\alpha_h(t)}{2}}B_h(t,Q^2,M_v^2)\left(\frac{1+\xi}{\xi}\right)^{\alpha_h(t)}.
\end{equation}

By using the known solution (\ref{solGPD}) for (\ref{short_form_A}) we write:
\begin{equation}\label{solGPD_s}
	 \text{GPD}_s(x,\xi,\dots)=\sin\left(\frac{\pi\alpha_s}{2}\right)\frac{B_s}{\pi}\frac{g_s(\xi)}{g_s(x)}\left(\frac{(1+\xi)(1-x)}{(1+x)(1-\xi)}\right)^{\frac{\alpha_s}{2}}\left(\frac{1+\xi}{\xi}\right)^{\alpha_s},
\end{equation}
\begin{equation}\label{solGPD_h}
	 \text{GPD}_h(x,\xi,\dots)=\sin\left(\frac{\pi\alpha_h}{2}\right)\frac{B_h}{\pi}\frac{g_h(\xi)}{g_h(x)}\left(\frac{(1+\xi)(1-x)}{(1+x)(1-\xi)}\right)^{\frac{\alpha_h}{2}}\left(\frac{1+\xi}{\xi}\right)^{\alpha_h},
\end{equation}
where $g_s(\xi,t,Q^2,M_v^2)$ and $g_h(\xi,t,Q^2,M_v^2)$ are two independent arbitrary real-valued functions that have only integer order zeros and/or singularities for $\xi\in[-1,1]$.

The resulting GPD will be the sum of (\ref{solGPD_s}) and (\ref{solGPD_h}).

\section{Conclusions}
We have suggested a method for deconvoluting general parton distributions beyond the cross-over line based on the assumed factorization properties of the GPD under the convolution integral. We test the method by using two simple, explicit Regge-pole based models of the DVCS amplitude. It is important that our scattering amplitudes, due to the presence of the Regge phases, are complex-valued functions. Otherwise, the phase can be recovered either by dispersion relation methods or from the interference between the DVCS amplitude with Bethe-Heitler processes, see \cite{Kresimir}. We hope that this method can be applied to other models of DVCS (or vector meson production (VMP)). Comparison with alternative models of DVCS (VMP) and the resulting GPDs is of great interest.

Although the models \cite{Capua} and \cite{Reggeometry}, used in the present paper, contain some $Q^2$ variation (introdeced phenomenologically), its value was fixed for simplicity at some $Q_0^2$ that can be interpreted as a fixed QCD running constant.

\section*{Acknowledgements}
%\small
 We thank V. Magas and F. Paccanoni for discussions. The work of L.J. was supported by the Program "Fundamental properties of matter under extreme conditions" of
 the National Academy of Sciences of Ukraine.

\end{document}